\documentclass[aps,pra,twocolumn]{revtex4}

\usepackage{graphicx}

\begin{document}


\title{ Ground state properties of a dilute homogeneous Bose gas of
hard disks in two dimensions
}
\author{F.Mazzanti$^1$, A.Polls$^2$, A. Fabrocini$^3$}
\address{ $^1$ Departament d'Electr\`onica, Enginyeria i Arquitectura
  La Salle, \\ 
  Pg. Bonanova 8, Universitat Ramon Llull, \\
  E-08022 Barcelona, Spain}
\address{ $^2$ Departament d'Estructura i Constituents de la 
  Mat\`eria, \\
  Diagonal 645, Universitat de Barcelona, \\
  E-08028 Barcelona, Spain}
\address{$^3$Dipartimento di Fisica "E.Fermi", Universit\`a di Pisa, 
  and INFN, Sezione di Pisa, \\ 
  Via Buonarroti,2 \\
  I-56100 Pisa, Italy}


\begin{abstract}
The energy and structure of a dilute hard-disks Bose gas are studied
in the framework of a variational many-body approach based on a
Jastrow correlated ground state wave function. The asymptotic
behaviors of the radial distribution function and the one-body density
matrix are analyzed after solving the Euler equation obtained by a
free minimization of the hypernetted chain energy functional. Our
results show important deviations from those of the available low
density expansions, already at gas parameter values $x\sim 0.001$. The
condensate fraction in 2D is also computed and found generally lower
than the 3D one at the same $x$.

\end{abstract}

\narrowtext

\maketitle

\section {Introduction}

The study of two dimensional (2D) quantum many--body systems is a
subject of considerable interest both from the theoretical and the
experimental points of view.  Atomic helium monolayers have attracted
much attention in these respects\cite{EK_helium2D}, and recently
Bose--Einstein condensation (BEC) in a strongly asymmetrical, quasi
two--dimensional trap has been achieved\cite{gorlitz01,rychtarik04}.
In a homogeneous 2D gas the one--body density matrix acquires a power
law decay below some low critical temperature, $T<T_c$,\cite{Tc} and,
at $T$=0, its $r\rightarrow \infty$ limit differs from zero. This
behavior defines a true condensate and BEC can occur in such a limit.
The modified density of states in confined geometry, as it is the case
for atoms in harmonic traps, makes BEC appear even at finite
temperature in 2D\cite{bagnato91,petrov00}.

In a recent paper\cite{mazzanti03}, referred as I hereafter, we have
examined the structure of the ground state of a 3D homogeneous gas of
bosons, interacting through both hard and soft core potentials. The
study was carried out to enlighten the role of the interaction induced
correlations along the density of the system and the possible
occurrence of {\em universality} in the dependence of different
properties on the gas parameter, $x=n a^3$ ($n$ being the particle
density and $a$ the $s$--wave scattering length).  The calculations
were performed within the Correlated Basis Functions (CBF) theory,
using optimized Jastrow correlated wave functions\cite{feenberg}. The
results were in excellent agreement with the existing Diffusion Monte
Carlo\cite{giorgini99} ones, obtained by the {\em exact} stochastic
solution of the many--body Schr\"odinger equation. Moreover, the
reliability (and limitations) of the analytical expansions in $x$ of
the energy per particle and of the condensate fraction were
clearly assessed \cite{lee57}.

In this work we use the same variational approach to study the $T=0$
ground state of a homogeneous gas of $N\rightarrow \infty$ hard--disks
(HD) of radius $a$.  We pay particular attention to the long--range
structure of the wave function, which is in turn intimately related to
the long--range order of the density matrix and to the condensate
fraction.

At present, there are not published DMC results for these systems in
the literature. However, based on the 3D case experience we believe
that the variational approach may provide a very accurate description
of the 2D ground state properties and serve as a test of validity for
the available low density expansions (LDE). This fact may be of
relevance also to the study of BEC in 2D harmonic traps, since LDE
are related to the Gross--Pitaevskii equation and its
modification\cite{fabrocini99} . LDE for the energy per particle in
terms of the 2D gas parameter, $x=n a^2$, have been derived by several
authors, starting from the leading order\cite{schick71}:
\begin{equation}
{{E_{LO}}\over N} ={4\pi x \over \mid\ln(x)\mid} \ , 
\label{E_LO}
\end{equation}
where the energy is in units of $\hbar^2/2 m a^2$.

An interesting question, that we do not address in this paper, is the
insurgence and the degree of universality in 2D. In the 3D Bose gas a
dependence in the energy on the shape of the potential appears already
at $x\sim 0.001$\cite{mazzanti03,giorgini99,fantoni02}, and breaks
down earlier for quantities other than the energy, as the short range
structure of the distribution function and the condensate fraction. A
similar study for the 2D gas would be of great interest by itself and
also in view of the analysis of the quasi 2D BEC experiments in terms
of the Gross--Pitaevskii equation.

In the case of strong interactions, as a hard--core potential,
correlation effects are crucial. Within CBF theory they are included
by means of a many--body correlation operator, ${\cal F}(1,2,\ldots ,
N)$, acting on the non--interacting ground state wave function,
$\Phi_0(1,2,\ldots , N)$,\cite{feenberg}:
\begin{equation}
\Psi_0(1,2,\ldots , N)=
{\cal F}(1,2,\ldots , N)
\Phi_0(1,2,\ldots , N) \, ,
\label{CBF}
\end{equation}
with $\Phi_0=1$ for homogeneous bosons.  ${\cal F}$ may be determined
according to the variational principle by minimizing the ground state
energy.

In I a Jastrow correlation operator was considered,
\begin{equation}
{\cal F} = \Psi_{J}(1,2,\ldots , N)=
\prod_{i<j} f(r_{ij})  \, ,
\label{Jastrow}
\end{equation}
where the two--body correlation function, $f(r_{ij})$, depends on the
interparticle distance, $r_{ij}=|{\bf r}_i-{\bf r}_j|$, vanishes for
$r_{ij}\le a$ and goes to unity at large distances. Expectation values
of operators on the Jastrow correlated wave function correspond to
multidimensional integrals (we consider only operators dependent on
the spatial coordinates) which can be either directly evaluated by
Monte Carlo type integrations (variational Monte Carlo, VMC) or
expanded in cluster diagrams.  The hypernetted chain (HNC) set of
integral equations allows for summing an infinite number of diagrams,
but it can be only approximately solved, since the class of the
$elementary$ diagrams is not summable in a closed way\cite{HNC}. The
approximation amounting to disregard the elementary diagrams (HNC/0)
is expected to be reliable in the low density regime, as shown in I.

In this paper we consider a system of $N$ spinless bosons of mass $m$
on a surface $S$, described by the hamiltonian
\begin{equation}
H = -{\hbar^2 \over 2m} \sum_{j=1}^N \nabla^2_j + 
\sum_{1=i<j}^N V(r_{ij})  \,
\label{hamil}
\end{equation}
where $V(r)$ is a two-body, symmetric hard--disk potential, 
\begin{equation}
V(r) = \left\{
\begin{array}{ll}
  \infty  & \,\,\,\,\, r<a \\
  0  & \,\,\,\,\, r> a \ ,
\end{array}
\right. 
\label{potHD}
\end{equation}
and the radius $a$ corresponds, for this potential, to the 2D
scattering length. We consider the system in the thermodynamic limit
($N$ and $S \rightarrow \infty$, keeping the density, $n = N/S$,
constant).

Minimization of the energy with respect to $f(r)$ provides the
$optimal$ Jastrow factor, which can be obtained through the solution
of the Euler--Lagrange (EL) equation\cite{cam,kro98}, $\delta E[f] /
\delta f=0$.  We adopt here the HNC/0 energy functional to solve the
EL equation.  The resulting correlation, $f_{EL}(r)$, shows a very
long range structure ($f_{EL}(r\rightarrow \infty) \rightarrow
1-\alpha /r$, with $\alpha$ constant) not easily accessible to VMC,
due to the limited size of the simulation box. As a consequence we
have also used a parametrized shorter range correlation factor,
suitable to be used in VMC and whose parameters are variationally
fixed. The comparison between the HNC/0 and VMC calculations with this
short range correlation provides a check of the accuracy of the
truncation in the cluster expansion.  Diffusion Monte Carlo is also
likely to suffer from the long range behavior of the wave function, as
these fine structures, related to collective effects, are hardly
distinguishable by numerical, finite size simulations.

 
The plan of the paper is as follows: the correlated variational theory
and its implementation (HNC and Euler equations) are briefly outlined
in Section II; an analysis of the EL asymptotic behaviors is presented
in Section III; Section IV presents the results for the HD model;
while summary and conclusions are given in Section V.

\section{Variational theory}

Given the Jastrow correlated wave function (\ref{Jastrow}), 
the energy per particle of the 2D homogeneous Bose gas is:  
\begin{equation}
{E\over N} = \frac {1}{2} n \int d{\bf r}_{12} \,g(r_{12}) 
\,\left [V(r_{12}) -
\frac{\hbar^2}{2m} \nabla^2 \ln f_2(r_{12}) \right ]  \, .
\label{energy}
\end{equation}

This expression simplifies for the HD potential (\ref{potHD}), 
since the radial distribution function (RDF),   
\begin{equation}
g(r_{12})=
{{N(N-1)}\over n^2}
{
{\int d{\bf r}_3 d{\bf r}_4 \ldots d{\bf r}_N 
\vert \Psi_0 \vert ^2}
\over
{\int d{\bf r}_1 d{\bf r}_2 \ldots d{\bf r}_N 
\vert \Psi_0 \vert ^2}
}\, ,
\label{RDF}
\end{equation}
vanishes inside the core of the potential and only the kinetic part
contributes to the energy.

The HNC equations provide a procedure to evaluate $g(r)$,
\begin{eqnarray}
g(r_{12}) & = & f_2^2(r_{12}) \, e^{N(r_{12})+E(r_{12})}
\nonumber \\ 
N(r_{12}) & = &  n  \int d {\bf r}_3 
[g(r_{13}) -1 ]
[g(r_{32}) -1 - N(r_{32})] \, ,
\label{HNC}
\end{eqnarray}
where $N(r)$ and $E(r)$ are functions representing the sum of the {\em
nodal} and the {\em elementary} diagrams, respectively \cite{HNC}.
The function $E(r)$ is an input to the HNC equations, which can be
solved once a choice for it has been done. The simplest possible
approximation corresponds to set $E(r)=0$ (HNC/0). This apparently
drastic truncation is, however, justified at low densities since the
elementary diagrams, due to their high connectivity, do not
appreciably contribute at the low densities relevant to BEC
experiments. Otherwise, the energy and the RDF can be stochastically
evaluated by Monte Carlo sampling of the corresponding many--body
integrals (\ref{RDF}).  The explicit calculation can be performed by
using the standard Metropolis algorithm \cite{metro} (see
Ref. \cite{guardiola} for a detailed description of Monte Carlo
methods).

Connected to the RDF is the static structure function (SSF), $S(k)$,
\begin{equation}
S(k) = 1 + n  \int d{\bf r} \,
e^{i {\bf k}\cdot {\bf r}} 
(g(r) -1)
\ ,
\label{SSF}
\end{equation}
often used in the analysis of (and extractable, within some
approximations, from) scattering experiments in condensed matter
physics (e.g. neutron scattering off liquid Helium\cite{helium}).

The Bose condensate is linked to the non--zero, long--range order of
the one--body density matrix (OBDM),
\begin{eqnarray}
n_1({\bf r}_1\!\!\!\! &,& \!\!\!\! {\bf r}_{1'}) =
n_1(r_{11'}) 
\\ [2mm]
& = & \!\! N { \int d{\bf r}_2 \cdots d{\bf r}_N 
\Psi_0(1,2, \ldots, N) 
\Psi_0(1',2, \ldots, N) 
 \over 
\int d{\bf r}_1 \cdots d{\bf r}_N 
\vert \Psi_0 \vert ^2 } \ .
\nonumber
\label{OBDM}
\end{eqnarray}
In fact, $n_1(r_{11'}\rightarrow \infty)/n = n_0$, where $n_0$ is the
condensate fraction. The depletion of $n_0$ with respect to unity is
an unmistakable indication of interparticle interactions (and, as a
consequence, of correlations). The Fourier transform of the OBDM
provides the momentum distribution (MD),
\begin{eqnarray}
n(k) \!\! & = & \!\! \int d{\bf r} \,e^{i{\bf k\cdot\bf r}}\,n_1(r) 
\\ 
& = & \!\!
(2\pi)^2 n\,n_0 \,\delta({\bf k})
+ \int d{\bf r} \,e^{i{\bf k\cdot\bf r}}
\left( n_1(r) - n\,n_0 \right) 
\nonumber
\label{MD}
\end{eqnarray}

As for the distribution function, the correlated OBDM can be computed
by using HNC theory\cite{HNC-ww}. In fact, $n_1(r)$ can be expressed
in terms of new nodal and elementary functions as
\begin{equation}
{{n_1(r)}\over n}  =  n_0 e^{N_{ww}(r)+E_{ww}(r)}
\label{rho1hnc}
\end{equation}
where $N_{ww}(r)$ is the solution of a generalized HNC
equation.
Again, setting $E_{ww}(r)=0$ these equations
can be solved in the HNC/0 approximation.

%

An appropriate choice of the correlation factor is essential for the
effectiveness of the variational approach. As stated in the
Introduction, the {\em best} choice is the one satisfying the
Euler--Lagrange equation,
\begin{equation}
{\delta E[g] \over \delta g(r)} = 0 \, ,
\label{Euler_g}
\end{equation}
which has been written in terms of the RDF rather than $f(r)$, since
in the HNC/0 scheme there is a one-to-one correspondence between these
two quantities. The correlation function can then be obtained by
inversion of the HNC equations. The EL equations are solved both in
configuration and momentum space, as discussed at length in I for the
3D hard--sphere case.  The theory can be straightfordwardly applied to
the 2D hard--disks gas and it is briefly outlined in the Appendix.

The correlation function produced by the solution of the EL equation
shows a long--range structure that is discussed in the next Section.
Finite size Monte Carlo techniques have difficulties to correctly deal
with this long range behavior and correlations healing to unity
inside the simulation box are used. In this respect, we have also
adopted a parametrized shorter--range correlation, $f_{SR}(r)$,
obtained by minimizing the energy computed at the two--body order of
the cluster expansion, $g_{2B}(r)=f^2(r)$, constrained by a
normalization condition
\begin{eqnarray}
{\delta \over \delta f(r)} \Bigg[
{1\over 2} \!\!\! & n & \!\!\!\!
\int d{\bf r}\,f^2(r)\left( -{\hbar^2
  \over 2m} \right) \nabla^2 \ln f(r) 
\nonumber \\ 
& & + \mu n \int d{\bf r}\left( 1- f^2(r) \right)
\Bigg] = 0
\label{min-heal}
\end{eqnarray}
which requires the use of a Lagrange multiplier $\mu$. The
minimization is performed under the {\em healing condition}
$f_{SR}(r\geq d)=1$, while $\mu$ is fixed by imposing
$f'(d)=f''(d)=0$.  The healing distance $d$ is taken as a variational
parameter to minimize the many--body energy.

For the particular case of the hard-core potential, the short 
range correlation, $f_{SR}(r)$, vanishes at $r\leq a$, while
\begin{equation}
f_{SR}(a\leq r\leq d) = 
{Y_0(\lambda a) J_0(\lambda r) - J_0(\lambda a) Y_0 (\lambda r)
\over Y_0(\lambda a) J_0(\lambda d) - J_0(\lambda a) Y_0 (\lambda d) }
\, ,
\label{f-SR}
\end{equation}
where $\lambda=-4m\mu/\hbar^2$ and $J_0$ and $Y_0$ are Bessel functions of 
the first and second kind, respectively. In the $d\rightarrow \infty$ limit, 
$f_{SR}$ coincides with the exact solution of the 2D zero energy Schr\"odinger 
equation, $f(r) \propto \ln(r)$~\cite{Miller}. 

\section{Asymptotic behaviors}

Applying to the 2D gas the sum-rule analysis of Ref.\cite{feenberg},
it can be shown that the SSF at low momenta has a linear dependence,
similar to the 3D gas:
\begin{equation}
S(k\rightarrow 0) \sim \frac {\hbar}{2 m c} k,
\label{asym1}
\end{equation}
where $c$ is the sound velocity in the medium. In terms of the RDF and
of the correlation function, it corresponds to a long range behavior,
different, however, from the 3D case where $(g_{3D}-1)\propto
1/r^{-4}$ and $(f_{3D}-1)\propto 1/r^{-2}$ at large $r$--values.

In fact, the 2D Fourier transform for a function, $h(r)$,  with circular 
symmetry can be written as:
\begin{equation}
\tilde h(k) = 2 \pi n \int_0^{\infty} dr  rf(r) J_0(kr) \, , 
\label{fourier}
\end{equation}
while $h(r)$ is obtained from $\tilde h(k)$ as
\begin{equation}
h(r) = \frac {1}{ 2 \pi n } \int_0^{\infty} dk  k \tilde h(k) J_0(kr) \, .  
\label{fourier1}
\end{equation}

Assuming that $ \tilde H(k)\equiv k \tilde h(k)$ is a well behaved
function at the origin, one finds
\begin{equation}
h(r \rightarrow \infty ) = \frac {\tilde H(0)}{r} - \frac {1}{2} \frac
{\tilde H^{II}(0)}{r^3} + \frac {3}{8} \frac {\tilde H^{IV}(0)}{r^5} + 
\cdots \, ,
\label{fourier2}
\end{equation}
where only the $k=0$ values of the function and of its even derivatives 
enter in the expansion.

The RDF is obtained from $S(k)$ through:
\begin{equation}
g(r) - 1 = \frac {1}{ 2 \pi n } \int_0^{\infty} dk~ k ~(S(k) -1 )
J_0(kr) \, ,
\end{equation}
so one readily obtains, from Eqs.~(\ref{asym1}) and~(\ref{fourier2}),
the asymptotic limit
\begin{equation}
g(r \rightarrow \infty ) = 1 - 
\frac {1}{2 \pi n } 
\frac {\hbar}{2  m c } \frac {1}{r^3}  \, . 
\label{LR-g}
\end{equation}

By inverting the HNC/0 equations (\ref{HNC}) one finds for the nodal
and correlation functions the following limits:
\begin{equation}
N(r \rightarrow \infty ) \rightarrow \frac{1}{2 \pi n} \frac {2m c}
{\hbar} \frac {1}{r} \, ,
\label{LR-N}
\end{equation}
and
\begin{equation}
f( r \rightarrow \infty ) \rightarrow 1 - \frac {1}{4 \pi n } 
 \frac {2mc}{\hbar} \frac {1}{r} \, ,
\label{LR-f}
\end{equation}
showing that in 2D, correlations have longer range than in 3D.

The long range structure of the OBDM is derived from the previous
expressions and the HNC equations. In fact, given the structure
(\ref{LR-f}) we obtain for $N_{\omega \omega}(r)$
\begin{equation}
N_{ww} (r\rightarrow \infty ) =\frac {1}{4 \pi n} \frac {m
c}{\hbar} \frac {1}{r} \, ,
\end {equation}
and for the OBDM
\begin{equation}
\frac {n_1 (r \rightarrow \infty )} {n} = 
n_0 + n_0 \frac {1}{4 \pi n} \frac
{m c}{\hbar} \frac {1}{r} \, .
\label{LR-OBDM}
\end{equation}

The momentum distribution has the same long-wavelength limit shown in
3D~\cite{gavoret64}, namely
\begin{equation}
\lim_{k \rightarrow 0} k n(k) = \frac {n_0}{2} \frac {mc}{\hbar} \, .
\end{equation}

\section{Results}

In this section we present and analyze results for the energy, radial
distribution function, static structure function and one--body density
matrix of the hard--disks gas.  We have used the optimized Jastrow
wave function obtained from the solution of the Euler-Lagrange
equations and the short-range correlation of Eq.~(\ref{f-SR}), mainly
to establish a comparison between the HNC/0 and the VMC approaches. In
the following, dimensionless quantities will be used: energies and
distances will be given in units of $\hbar^2/2 m a^2$ and $a$
respectively.

Several corrections to $E_{LO}$ have been proposed in the literature. 
Kolomeisky and Straley~\cite{kolo92} used renormalization group 
techniques to study the ground state of dilute Bose systems
as a function of the space dimensionality. They found a general
expression, valid for strong interactions when $x\to 0$, that simplifies
for hard disks to
\begin{equation}
{E_{KS}\over {E_{LO}}} 
 = -{\mid\ln(x)\mid \over \ln(4\pi x)}
\left[ 1 - { \ln \left(-\ln(4\pi x) \right) \over \ln(4\pi x)} \right] \ .
\label{E_kolo}
\end{equation}

Cherny and Shanenko~\cite{cherny01} derived an alternative expansion,
\begin{equation}
{E_{CS}^{(u)}\over {E_{LO}}} 
 = \mid\ln(x)\mid 
\left[ u + {u^2 \over 2} + \cdots \right] \, ,
\label{E_chernu}
\end{equation} 
 in the parameter $u$ satisfying the equations: 
\begin{equation}
u=\delta(1+u\ln u)
\,\,\,\,\, , \,\,\,\,\,
\delta = -{1\over \ln(\pi x) + 2\gamma} \, ,
\label{ener-cher-eqs}
\end{equation}
where $\gamma=0.577\ldots$ is the Euler constant. These authors also
gave an expansion of $u$ in terms of $\delta$, allowing to write the
series (\ref{E_chernu}) in the form:
\begin{equation}
{E_{CS}^{(\delta)}\over {E_{LO}}} 
\! = \mid\!\ln(x)\!\mid\!\! 
\left[ 
\delta + \delta^2 \ln\delta + {\delta^2 \over 2} + 
\delta^3\ln^2\delta + 2\delta^3 \ln\delta + \cdots \right] \ .
\label{E_chernd}
\end{equation}

Table~\ref{tab-1} reports the energy per particle as a function of $x$
in the EL approximation (EL/HNC), the variational Monte Carlo
calculation starting from $f_{SR}(r)$ (SR/VMC) and the HNC approach
with the same correlation function (SR/HNC). We remind that the HNC/0
approximation is used everywhere but in the VMC.  The results of the
$x$--expansions previously discussed are also reported.

The comparison between SR/VMC and SR/HNC shows that the influence of
the missing elementary diagrams on the energy is less than $1\%$,
except at the highest value $x=0.1$ ($\sim 2.3\%$). This gives us
confidence that the variational principle is mostly satisfied within
our HNC/0 calculations, providing the hierarchy: $E_{exact}\leq E_{EL}
\leq E_{SR}$. Only at $x=0.1$ these inequalities do not numerically
hold. The cases $x=10^{-5}$ and $x=5\cdot 10^{-2}$ can be considered to
fulfill the inequalities if we take into account the numerical accuracy
associated to the calculation at these quantities.  If we estimate the
contribution of the elementary diagrams on $E_{EL}$ by scaling it by
the ratio $E^{VMC}_{SR}/E^{HNC}_{SR}$, we obtain
$E_{EL}/N(x=0.1)=0.9075$, restoring all the inequalities.

Lieb~\cite{lieb63} pointed out that a lower bound to the exact energy
is given by
\begin{equation}
E_{low} = E_{LO} 
\left[ 1 - 
{\mathcal O}\left(\mid \ln(x) \mid^{-1/5}\right) \right] \ ,
\label{ener-bound}
\end{equation}
and that $E_{exact}/E_{LO}\to 1$ when $x\to 0$. 
Both the variational energies (EL and SR) comply with condition 
(\ref{ener-bound}) and seem to tend to $E_{exact}$ when $x$ goes to zero. 

Table~(\ref{tab-1}) gives also the healing distance, $d$, of the SR
correlation in units of $a$. $d$ increases when $x\to 0$, the Lagrange
multiplier decreases and the energy goes to zero.  Therefore,
$f_{SR}(r)_{x\to 0}$ can be approximated by its $\lambda=0$ limit,
which coincides with the zero energy limit of the two--body
Scr\"odinger equation,
\begin{equation}
f_{SR}(r)_{\lambda\to 0} \to {\ln(r) \over \ln(d)} \ .
\label{limit-heal}
\end{equation}

\begin{figure}
\begin{center}
\includegraphics*[width=0.9\textwidth]{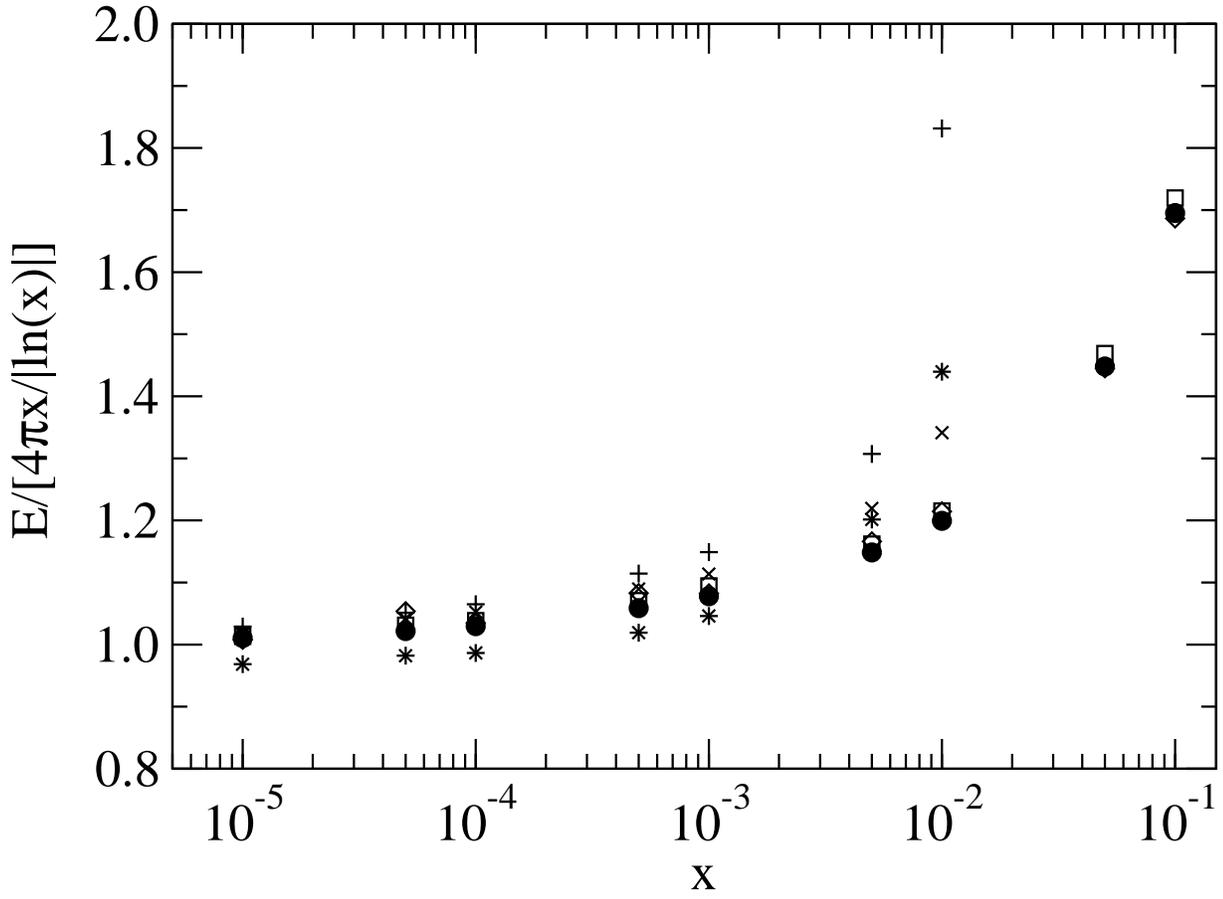}
\end{center}
\caption{Scaled energy per particle of the HD gas as a function of
  $x$. Solid circles: EL/HNC;
  open squares: SR/HNC; 
  open diamonds: SR/VMC;
  stars, pluses and crosses: low density expansions, 
  $KS$, $CS^{(u)}$ and $CS^{(\delta)}$, respectively. Notice that 
  diamonds are hardly distinguished from the solid circles.
}
\label{fig-ener}
\end{figure}

These results are also shown in Figure (\ref{fig-ener}), where the
variational scaled energies per particle (EL, SR/VMC and SR/HNC) are
compared with the $E_{KS}$ and $E_{CS}$ estimates.  All energies have
been divided by $E_{LO}$ in order to stress the deviations from the
low--density limit.  The limit is approached by $E_{EL}$ and $E_{SR}$
from above when $x$ decreases, although it has not been yet fully
reached at $x=10^{-5}$. The differences between the variational and
the low density energies are still visible, even in the density range
relevant to BEC experiments~\cite{rychtarik04}.

$E_{KS}$ does not satisfy the $x=0$ lower bound (\ref{ener-bound})
and, starting from $x\approx 0.005$, becomes higher than the
variational upper bound provided by $E_{EL}$ and $E_{SR}$.  $E_{CS}$
satisfies the low density limit, but lies above the variational upper
bounds at any value of $x$.  $E_{CS}^{(u)}$ is always larger than
$E_{CS}^{(\delta)}$ and the difference increases drastically along
$x$. Notice that Eq.~(\ref{ener-cher-eqs}) does not have solution at
$x\geq 0.0369$.

The EL optimization procedure does not significantly affect the 
energies obtained with $f_{SR}(r)$, since the
energy is dominated by the short range structure of the HD potential, 
requiring $g(r)$ to vanish inside the core. The effects of the 
long range structure of the EL correlation  are, instead, clearly 
evident in the behavior of the radial distribution function.

\begin{figure}
\begin{center}
\includegraphics*[width=0.9\textwidth]{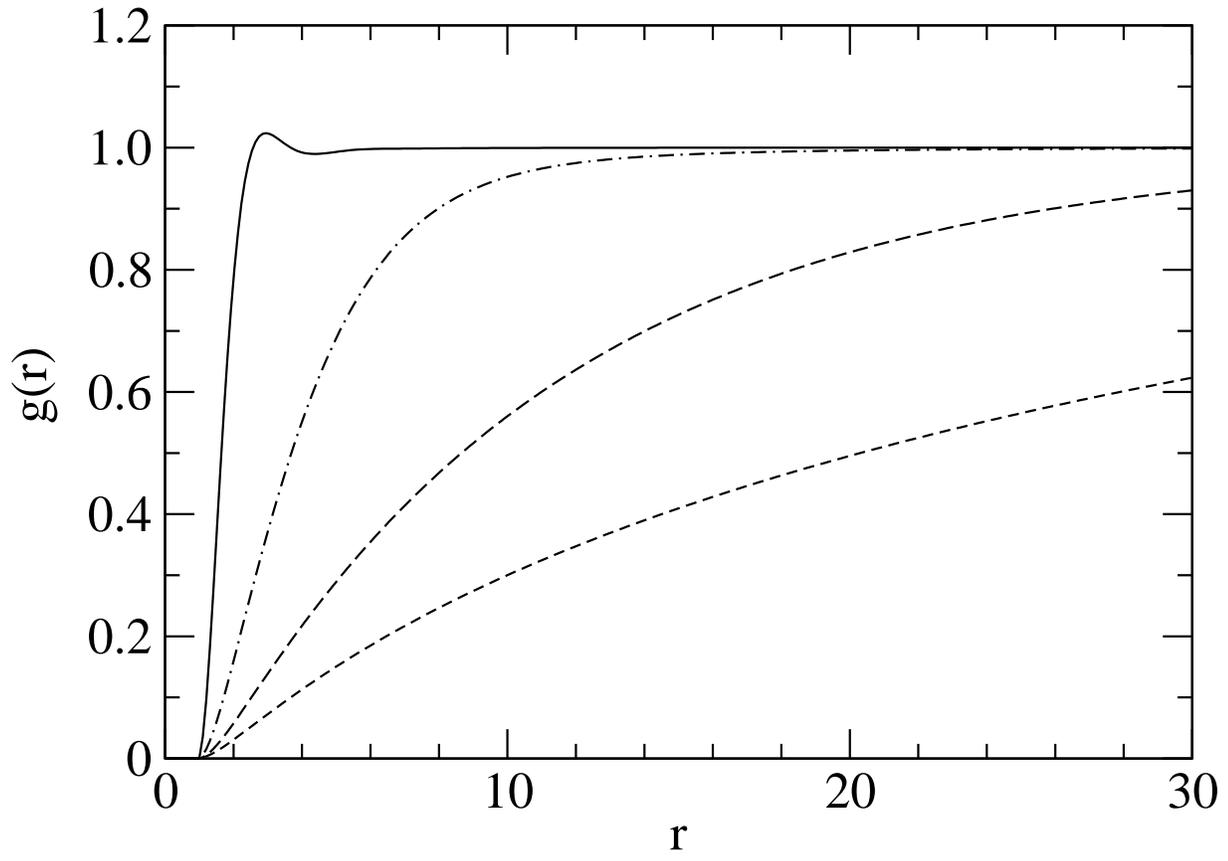}
\end{center}
\caption{EL radial distribution functions $g(r)$ for 2D hard disks at
 several values of $x$.  Solid, dot-dashed, long-dashed and
 short-dashed lines stand for $x=0.1,0.01,0.001$ and $0.0001$,
 respectively.
}
\label{fig-gr}
\end{figure}

The EL RDF is shown in Figure (\ref{fig-gr}) for different values of $x$. 
At low $x$, $g_{EL}(r)$ is a monotonically increasing
function of the distance. However, it develops a local maximum close
to the core radius at densities $x>0.01$. 
This is a genuine many--body effect induced by the strong correlations 
at high density. The same behavior was found in I for the 3D Bose gas. 
As expected, the correlation hole is more 
pronounced at larger densities. 

\begin{figure}
\begin{center}
\includegraphics*[width=0.9\textwidth]{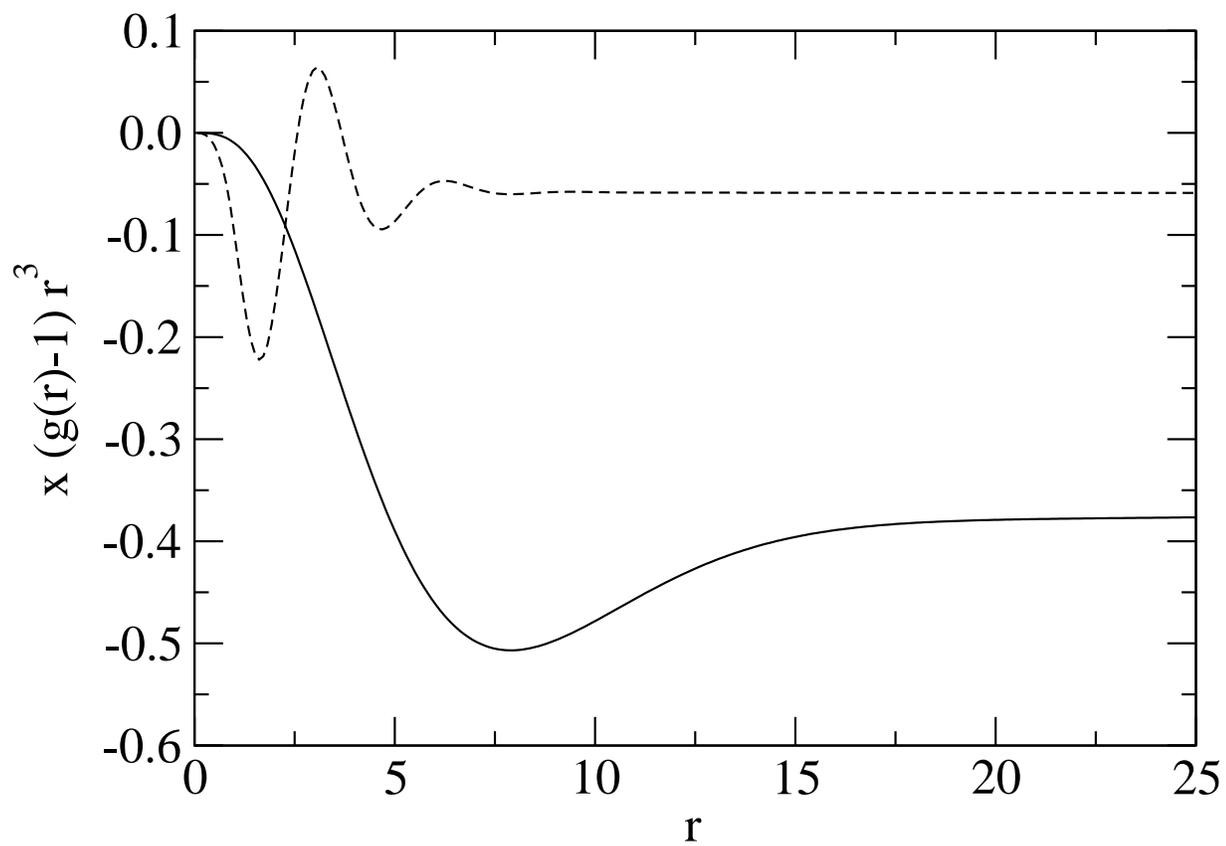}
\end{center}
\caption{Long range  structure of the EL $g(r)$ at $x=0.1$ (dashed 
 line)  and $x=0.01$ (solid line). 
 }
\label{fig-long-gr}
\end{figure}

\begin{figure}
\begin{center}
\includegraphics*[width=0.9\textwidth]{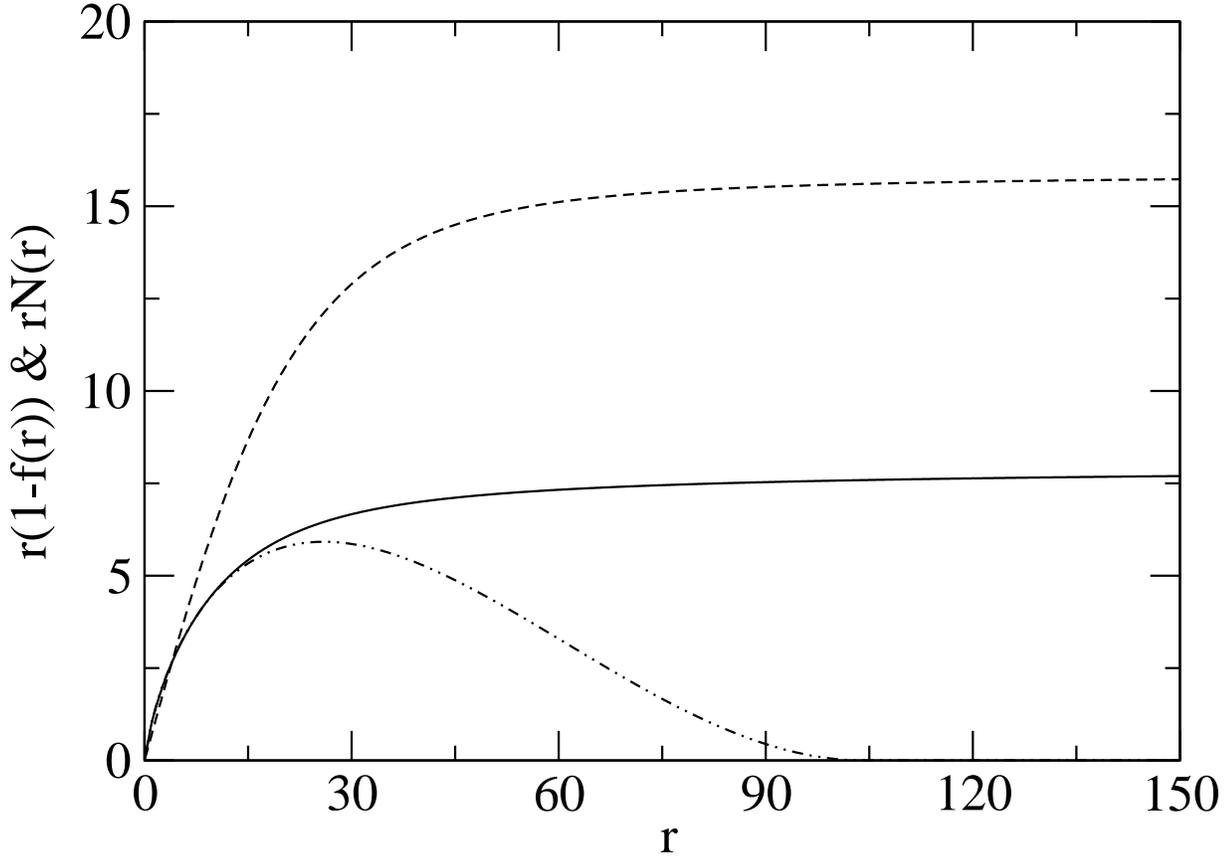}
\end{center}
\caption{EL  correlation (solid line),  SR correlation 
(dot-dot-dashed line) and nodal (dashed line) 
   functions at $x=10^{-3}$. 
  }
\label{fig-corr}
\end{figure}

The long range limit of $g_{EL}(r)$ is shown in Figure
(\ref{fig-long-gr}) at $x=0.01$ and $x=0.1$. The quantity shown is
$x\, r^3(g(r)-1)$ whose dimensionless asymptotic limit is:
\begin{equation}
x   r^3 [ g( r\to\infty) - 1 ] = - {1\over 2\pi c} \, .
\label{asymp-gr}
\end{equation}
This ratio is smaller at $x=0.1$, implying that the sound velocity
increases with $x$, as expected. Consistent with the previous figure,
the asymptotic limit is reached faster at larger densities. Figure
(\ref{fig-corr}) gives the EL correlation and the nodal function,
$N(r)$, at $x=0.001$.  We show $[x\, r (1 - f_{EL}( r))]$ and $[x\,r
N( r)]$ to enlighten the asymptotic limits (\ref{LR-f}) and
(\ref{LR-N}) whose dimensionless values are $c/4 \pi$ and $c/ 2 \pi$,
respectively. The fact that one limit is twice the other is
clearly appreciated in the figure.
Notice also that due to the chain process
implied by the HNC scheme, $N(r)$ merges to the $1/r$ law at larger
distances then $f(r)$.  To illustrate the different asymptotic
behaviors, we also show $[x\, r (1 - f_{SR}( r))]$ which goes quickly
to zero.

\begin{figure}
\begin{center}
\includegraphics*[width=0.9\textwidth]{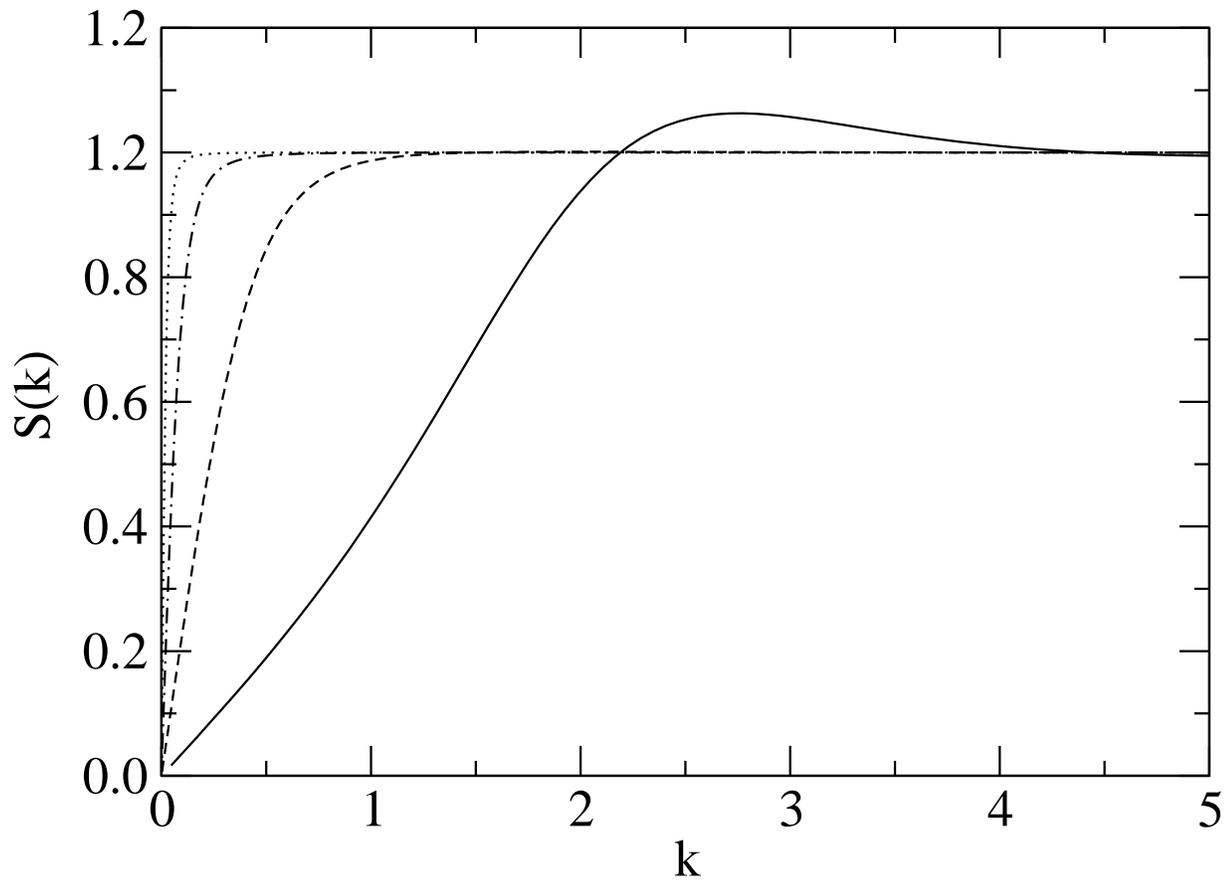}
\end{center}
\caption{
EL  static structure factor, $S(k)$, at several 
values of $x$. Solid line: $x=0.1$, dashed line: $x=0.01$, dot-dashed line: 
$x=0.001$, and dotted line: $ x=0.0001$.  
}
\label{fig-SSF}
\end{figure}

The EL static structure function, $S(k)$, is shown in Figure
(\ref{fig-SSF}).  At low densities the SSF reaches the asymptotic
value, $S(k)\to 1$, already at $k\sim 1$. As in the RDF case, the
overshooting of the SSF at the highest density, $x=0.1$, is a
consequence of the correlations. 
The linear regime of $S(k)$ around the origin is appreciable, although
the calculation of the ratio $S(k)/k$ shows 
deviations from a constant value already at low $k$.

\begin{figure}
\begin{center}
\includegraphics*[width=0.9\textwidth]{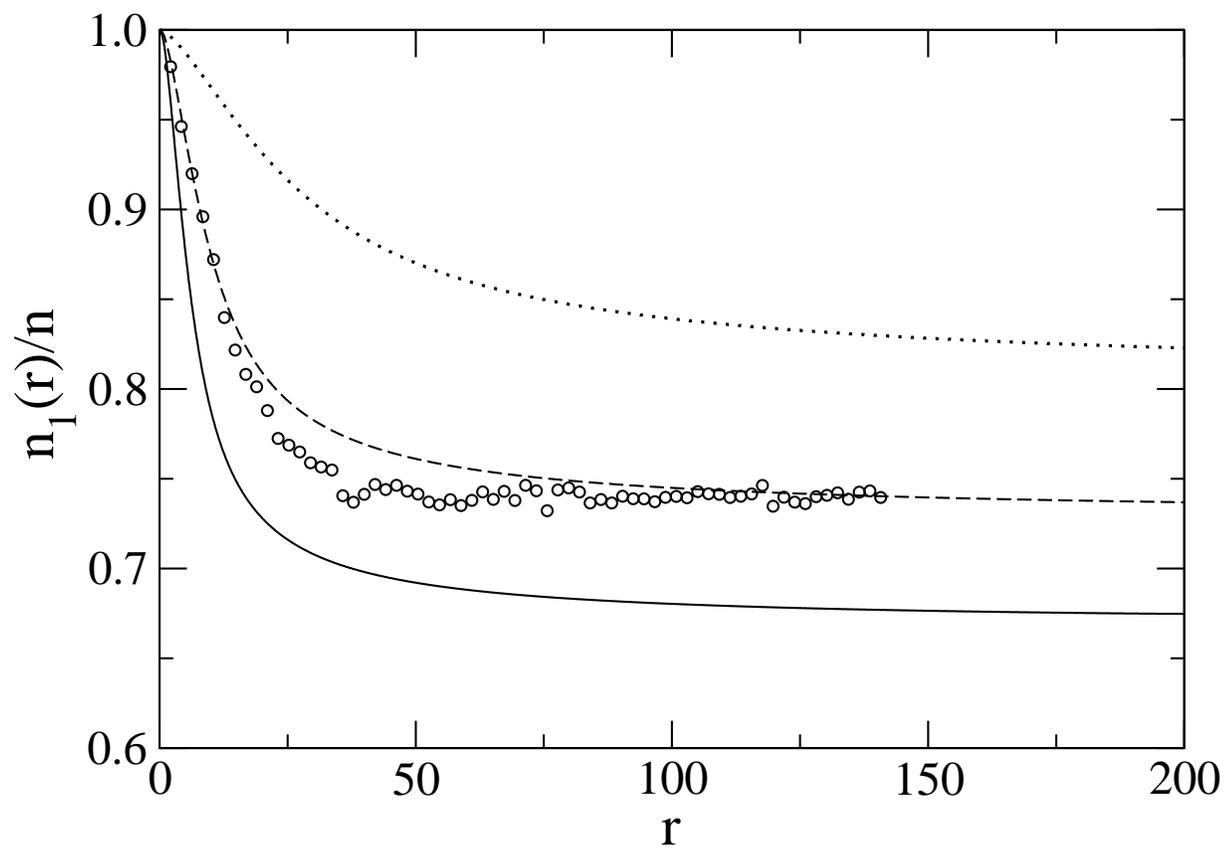}
\end{center}
\caption{
EL  one--body density matrices at several values of $x$.
Solid line: $x=0.01$, dashed line: $x=0.005$, and 
dotted line: $x=0.001$. Open circles, SR/VMC results at $x=0.005$.
}
\label{fig-OBDM}
\end{figure}

\begin{figure}
\begin{center}
\includegraphics*[width=0.9\textwidth]{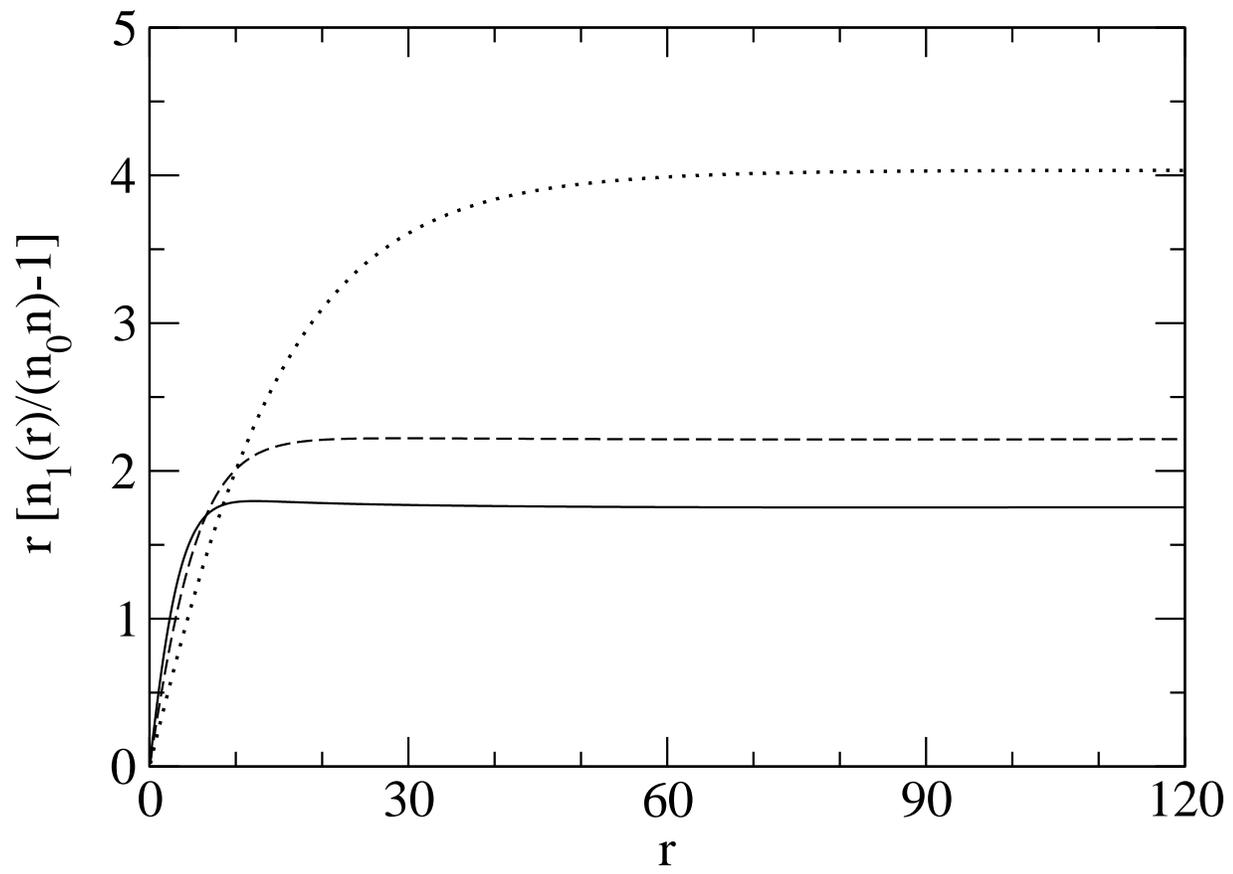}
\end{center}
\caption{
Long range behavior of the EL one--body density matrix 
at several values of $x$. Solid line: $x=0.01$,
 dashed line: $x=0.005$, and dotted line: $x=0.001$. 
}
\label{fig-long-OBDM}
\end{figure}

\begin{figure}
\begin{center}
\includegraphics*[width=0.9\textwidth]{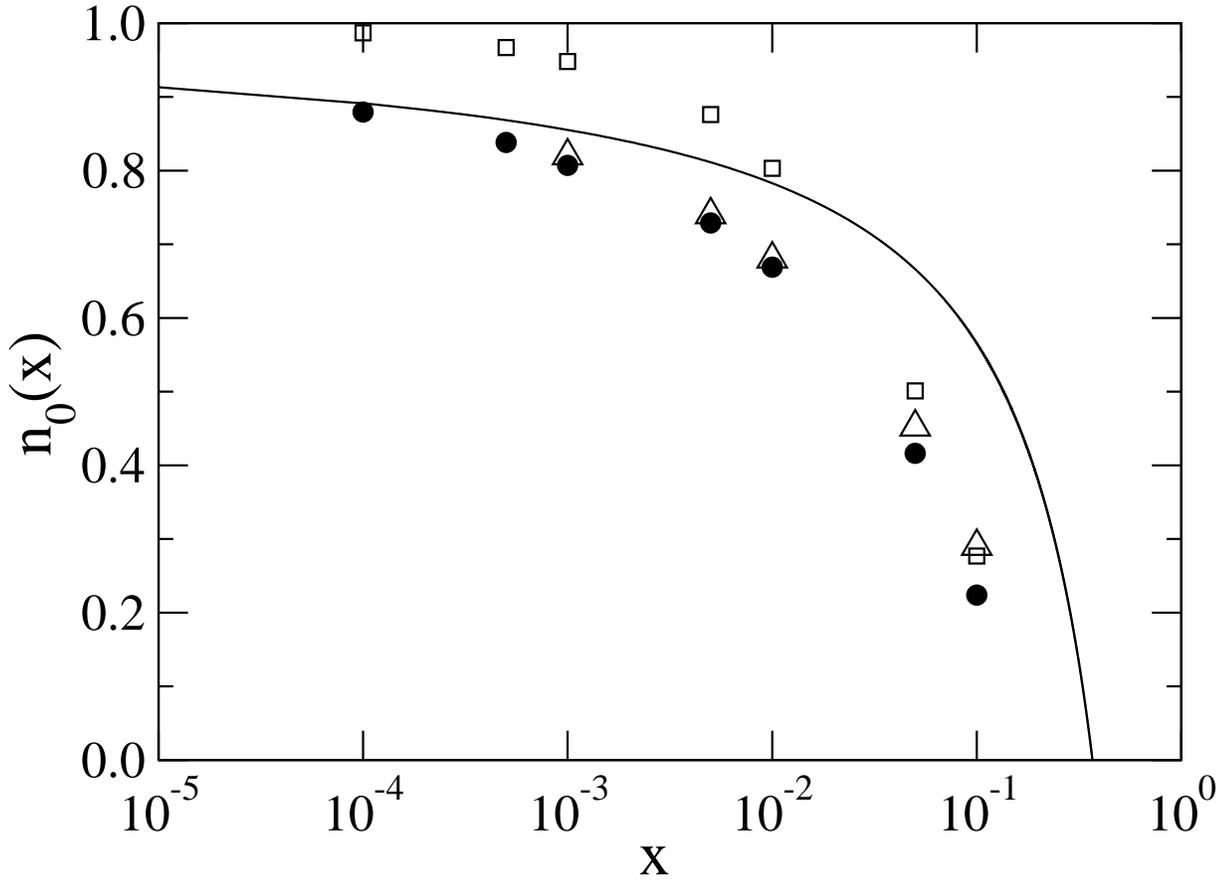}
\end{center}
\caption{Condensate fraction as a function of $x$. Black circles,
triangles and solid line correspond to the EL/HNC, SR/VMC and
low-density expansion results, respectively. Open squares stand for
the corresponding 3D values taken from I.}
\label{fig-n0}
\end{figure}

In figure~(\ref{fig-OBDM}) we plot the one--body density matrix,
$n_1(r)$, in the EL approach for $x=0.01$, $0.005$ and $0.001$. The
asymptotic limit of $n_1(r)$ defines the value of the condensate
fraction, which decreases when the gas parameter increases.  The
asymptotic value is reached faster when $x$ increases. The detailed
long-range behavior (\ref{LR-OBDM}) is presented in Figure
(\ref{fig-long-OBDM}) by showing the quantity $r[n_1( r)/(n_0n)-1]$,
whose dimensionless asymptotic value is $c/(8 \pi x)$. Even if the
speed of sound increases with $x$, the value of this limit is
dominated by the presence of the gas parameter in the denominator and
the overall quantity increases when $x$ decreases.  Finally, the EL
and SR/VMC condensate fractions, $n_0(x)$, are reported in Figure
(\ref{fig-n0}). The Figure also contains the low--density
prediction~\cite{schick71},
\begin{equation}
n_0^{LD}= 1 + {1\over\ln(x)} \ .
\label{n0_LD}
\end{equation}
$n_0^{LD}$ appears to sensibly overestimate the condensate fraction.
The EL and SR/VMC condensate fractions are very similar except for the
largest value of $x$ reported in the figure, where the contribution of
the elementary diagrams could be important.  This fact indicates that
the value of $n_0$ is not very much affected by the inclusion of a
long-range structure into the correlation function. However, the use
of $f_{EL}(r)$ is crucial to approach this value in a proper way, that
is, to satisfy Eq.~(\ref{LR-OBDM}).  Also reported in the figure is
the condensate fraction of the 3D system of hard spheres, taken from
I.  At fixed $x$, the 2D condensate fraction is smaller than the 3D
one, indicating that correlations in the 2D system are {\em stronger},
thus promoting more particles outside the zero-momentum state.

\section{Summary and Conclusions}

In this work we have analyzed the energy and structure of a
homogeneous gas of bosons in two dimensions interacting via a hard
disk potential whose core radius equals its corresponding 2D
scattering length.  We have adopted a variational many-body approach,
based on a Jastrow correlated ground state wave function. The
expectation values have been computed both in the framework of the
hypernetted chain theory (within the HNC/0 approximation) and with the
variational Monte Carlo method. Two types of correlation functions
have been used: {\em (i)} a long range one, obtained by the free
minimization of the HNC/0 ground state energy and preserving the
correct asymptotic behaviors of the wave function; {\em (ii)} a short
range one, to be used in the Monte Carlo sampling and providing a
check of the accuracy of the cluster expansion.

By comparing with the VMC results, the accuracy of the HNC/0 energies
are better than 1$\%$, except at the highest density, $x=0.1$,
where the error is still less than 3$\%$. The EL minimization lowers
the energy with respect to the SR correlation by $\sim 1.5\%$. We do
not expect further large reductions from a complete DMC
calculation. The low density expansions start to severely deviate from
the variational results already at $x\sim 0.001$, and the most
accurate of them appears to be the Cherny and Shanenko expansion in
terms of the parameter $\delta$. However, their use for estimating
corrections to the 2D Gross--Pitaevskii equation, especially in the
large gas parameter regime, seems questionable.

Finally, the condensate fractions lies well below the values predicted
by the low density theories and also below the results for the three
dimensional gas of hard spheres at the same gas parameter.

We conclude that the variational theory is a powerful and reliable
tool to study dilute systems, also in 2D.
Moreover, the 
homogeneous gas results may be used in a 
local density type approximation~\cite{fabrocini99} to analyze 
bosons in two dimensional harmonic traps.

\section{Appendix}

In this appendix we discuss the EL equations for an interacting
many--body system in 2D. As in the 3D case, the solution to the
optimization equation,
\begin{equation}
{\delta E[g] \over \delta g(r)} = 0 \, ,
\label{app-ELg}
\end{equation}
can be obtained in momentum space, yielding 
\begin{equation}
S(k) = {t(k) \over \sqrt{t^2(k) + 2 t(k) V_{ph}(k)}} \ ,
\label{app-Sk}
\end{equation}
where $t(k)=\hbar^2 k^2/2m$ is the free--particle energy spectrum and
$V_{ph}(k)$ is the particle--hole interaction. The latter can be
expressed in configuration space and reads, disregarding the
contribution of elementary diagramas
\begin{equation}
V_{ph}(r) = g(r) V(r) + {\hbar^2 \over m}
\left|\nabla\sqrt{g(r)}\right|^2 + \left[ g(r)-1 \right] \omega_I(r) \ ,
\label{app-Vphr}
\end{equation}
in terms of the induced interaction $\omega_I(r)$. In momentum space,
we have 
\begin{equation}
\omega_I(k) = -{1\over 2} t(k) { \left( 2 S(k)+1 \right) \left( S(k)-1
  \right) \over S(k) } \ .
\label{app-wI}
\end{equation}

Eqs.~(\ref{app-Sk}) to~(\ref{app-wI}) are to be solved
simultaneously. This can be done starting from a suitable choice for
$g(r)$, performing its FT to get $S(k)$, evaluating $\omega_I(k)$ and
$V_{ph}(r)$, and then deriving a new $S(k)$ with the help of
Eq.~(\ref{app-Sk}). This procedure is iterated until the difference
between two consecutive iterations is as small as desired.

Up to this point there are no formal differences between the 2D and 3D
cases.  The main deviation is the way in which the Fourier transforms
are carried out. In 2D and for a general function, $f(r)$, the FT and
its inverse read
\begin{eqnarray}
f(k) & = & 2\pi n \int_0^\infty dr\, r f(r)\,J_0(kr) \\
f(r) & = & {1\over 2\pi n} \int_0^\infty dk\, k f(k)\,J_0(kr)
\label{app-FTs}
\end{eqnarray}
where $n$ is the (constant) density and $J_0(x)$ the zero order Bessel
function of the first kind. One way to implement these transformations
is to use a finite box in configuration and momentum spaces of length
$L$ and $K$, respectively. The additional conditions $f(r=L)=0$ and
$f(k=K)=0$ lead to a discretized set of coordinates and momenta,
related to the zeros $\lambda_j$ of $J_0(x)$ through the relations
\begin{equation}
k_j = {\lambda_j \over L} 
\,\,\,\,\,\, , \,\,\,\,\,\,
r_\alpha = L{\lambda_\alpha \over \lambda_N}
\label{app-grids}
\end{equation}
with $j,\alpha=1, 2, \ldots, N$, $N$ being the total number of points
in the grids. A Gauss integration rule based on series expansion
in Bessel functions and the orthogonality relation,
\begin{equation}
\int_0^L dr\,r J_0(k_i r) J_0(k_j r) = {2\over L^2 J_1^2(k_j
L)}\,\delta_{ij} \, ,
\label{app-ortho}
\end{equation}
can then be built, leading to
\begin{eqnarray}
\int_0^L dr\, r f(r) J_o(k_j r) & = & \sum_{\alpha=1}^N 
\omega_\alpha J_0(k_j r_\alpha) f(r_\alpha) \, , \nonumber \\
\int_0^K dk\, k f(k) J_0(k r_\alpha) & = & 
\sum_{j=1}^N \omega_j J_0(k_j r_\alpha) f(k_j) \ ,
\label{app-gauss}
\end{eqnarray}
whith the integration weights 
\begin{equation}
\omega_\alpha = { 2L^2 \over \lambda^2_N J_1^2(\lambda_\alpha) }
\,\,\,\,\,\, , \,\,\,\,\,\,
\omega_j = { 2 \over L^2 J_1^2(\lambda_j) } \ .
\label{app-weights}
\end{equation}

Eqs.~(\ref{app-gauss}) turn integrals into algebraic products that can
be carried out numerically in a neat and fast way using available
standard libraries.

\begin{acknowledgments}

The authors are grateful to Profs. E. Krotscheck, S. Giorgini and
J. Boronat, and to S. Pilato, for useful discussions.  This work has
been partially supported by Grant No. BFM2002-01868 from DGI (Spain),
Grant No. 2001SGR-00064 from the Generalitat de Catalunya, and by the
Italian MIUR through the {\it PRIN: Fisica Teorica del Nucleo Atomico
e dei Sistemi a Molti Corpi}.

\end{acknowledgments}




\begin{widetext}

\begin{table}[!t]
\begin{center}
{
\begin{tabular}{|c|c|c|c|c|c|c|c|c|c|}
\hline \hline 
$x$ & ${E}_{EL}/N$ & ${E}^{VMC}_{SR}/N$  & $\epsilon_{VMC}$ &
${E}^{HNC}_{SR}/N$ & $d$ & ${E}_{LO}/N$  & ${E}_{KS}/N$ & 
${E}_{CS}^{(u)}/N$ & ${E}_{CS}^{(\delta)}/N$ 
\\ \hline
$10^{-5}$            & $1.103\cdot 10^{-5}$ & $1.100\cdot 10^{-5}$   & 
$2.47\cdot10^{-7}$   &
$1.106\cdot 10^{-5}$ & $1135.52$ &
$1.091\cdot 10^{-5}$ & $1.057\cdot 10^{-5}$ & $1.123\cdot 10^{-5}$ &
$1.117\cdot 10^{-5}$ \\
$5\cdot 10^{-5}$     & $6.482\cdot 10^{-5}$ & $6.685 \cdot 10^{-5}$   & 
$2.44\cdot10^{-6}$   &
$6.542\cdot 10^{-5}$ & $495.70$ & $6.344\cdot 10^{-5}$ &
$6.231\cdot 10^{-5}$ & $6.670\cdot 10^{-5}$ & $6.610\cdot 10^{-5}$ \\
$10^{-4}$            & $1.405\cdot 10^{-4}$ & $1.415\cdot 10^{-4}$ & 
$2.20\cdot10^{-6}$   &
$1.417\cdot 10^{-4}$ & $344.15$ & $1.364\cdot 10^{-4}$ & 
$1.346\cdot 10^{-4}$ & $1.453\cdot 10^{-4}$ & $1.436\cdot 10^{-4}$ \\
$5\cdot 10^{-4}$     & $8.752\cdot 10^{-4}$ & $8.952 \cdot 10^{-4}$  &
$2.03\cdot10^{-5}$   &
$8.851\cdot 10^{-4}$ & $153.49$ & $8.266\cdot 10^{-4}$ &
$8.425\cdot 10^{-4}$ & $9.212\cdot 10^{-4}$ & $9.007\cdot 10^{-4}$ \\
$10^{-3}$            & $1.961\cdot 10^{-3}$ & $1.979\cdot 10^{-3}$ & 
$2.87\cdot10^{-5}$   &
$1.991\cdot 10^{-3}$ & $104.54$ & $1.819\cdot 10^{-3}$ & 
$1.903\cdot 10^{-3}$ & $2.090\cdot 10^{-3}$ & $2.026\cdot 10^{-3}$ \\
$5\cdot 10^{-3}$     & $1.362\cdot 10^{-2}$ & $1.383\cdot 10^{-2}$ &    
$8.80\cdot10^{-5}$   &
$1.378\cdot 10^{-2}$ & $46.33$  & $1.186\cdot 10^{-2}$ & 
$1.435\cdot 10^{-2}$ & $1.550\cdot 10^{-2}$ & $1.446\cdot 10^{-2}$ \\
$10^{-2}$            & $3.273\cdot 10^{-2}$ & $3.303\cdot 10^{-2}$ & 
$1.30\cdot10^{-4}$   &
$3.316\cdot 10^{-2}$ & $33.41$  & $2.729\cdot 10^{-2}$ &
$3.928\cdot 10^{-2}$ & $4.998\cdot 10^{-2}$ & $3.660\cdot 10^{-2}$ \\
$5\cdot 10^{-2}$     & $0.3037$             &  $0.3031$            & 
$0.00089$            &
$0.3081$             & $17.82$              & $0.2097$ &
$3.5818$             & NA                   & NA \\
$0.1$                & $0.9252$             & $0.9204$ &
$0.003$              &
$0.9384$             & $15.66$              & $0.5458$ &
NA                   & NA                   & NA \\
\hline \hline
\end{tabular}
}
\end{center}
\caption{ Energy per particle for the hard disk model, as a function
of $x$.  $E_{EL}$ is obtained by solving the HNC/0 EL equation,
$E_{VMC}$ and $E_{SR}$ are the VMC and HNC/0 energies with the SR
correlation having a healing distance $d$, in units of the scattering
length. $\epsilon_{VMC}$ is the statistical error in the variational
Monte Carlo calculation.  The last four columns give the energies in
different low density expansions (see text).  }
\label{tab-1}
\end{table}

\end{widetext}

\pagebreak


\pagebreak

\pagebreak

\pagebreak

\pagebreak

\pagebreak

\pagebreak

\pagebreak

\pagebreak

 \end{document}